\documentclass[journal=jpclcd,manuscript=article]{achemso}
\usepackage{chemformula} 
\usepackage[T1]{fontenc} 
\usepackage{mathtools}
\usepackage{amsthm}
\usepackage{amssymb}
\usepackage{bm}
\usepackage[version=3]{mhchem}
\usepackage{braket}
\usepackage{relsize}
\usepackage{enumitem}
\usepackage[vlined]{algorithm2e}
\usepackage{algpseudocode}
\usepackage{graphicx}
\usepackage{footnote}
\usepackage{multirow}
\usepackage{tabularx}
\usepackage{booktabs}
\usepackage{makecell}
\usepackage{threeparttable}
\usepackage{color}
\usepackage{breakurl}

\newcolumntype{C}{>{\centering\arraybackslash}X}


\title{A Multi-Resolution 3D-DenseNet for Chemical Shift Prediction in NMR Crystallography}

\author{Shuai Liu}
\author{Jie Li}
\author{Kochise C. Bennett}
\author{Brad Ganoe}
\author{Tim Stauch} 
\author{Martin Head-Gordon}
\affiliation{
Pitzer Center for Theoretical Chemistry, Department of Chemistry,  University of California, Berkeley CA 94720
}

\author{Alexander Hexemer}
\affiliation{ 
Advanced Light Source, Lawrence Berkeley National Laboratory, Berkeley CA 94720}

\author{Daniela Ushizima}
\affiliation{ 
Computational Research Division, Center for Advanced Mathematics for Energy Research Applications (CAMERA), Lawrence Berkeley National Laboratory, Berkeley CA 94720}

\author{Teresa Head-Gordon}
\affiliation{
Pitzer Center for Theoretical Chemistry, Department of Chemistry, Department of Bioengineering, Department of Chemical and Biomolecular Engineering, University of California, Berkeley CA 94720}
\email{thg@berkeley.edu}
\begin{document}

\begin{tocentry}
\centering
\includegraphics[width=\linewidth]{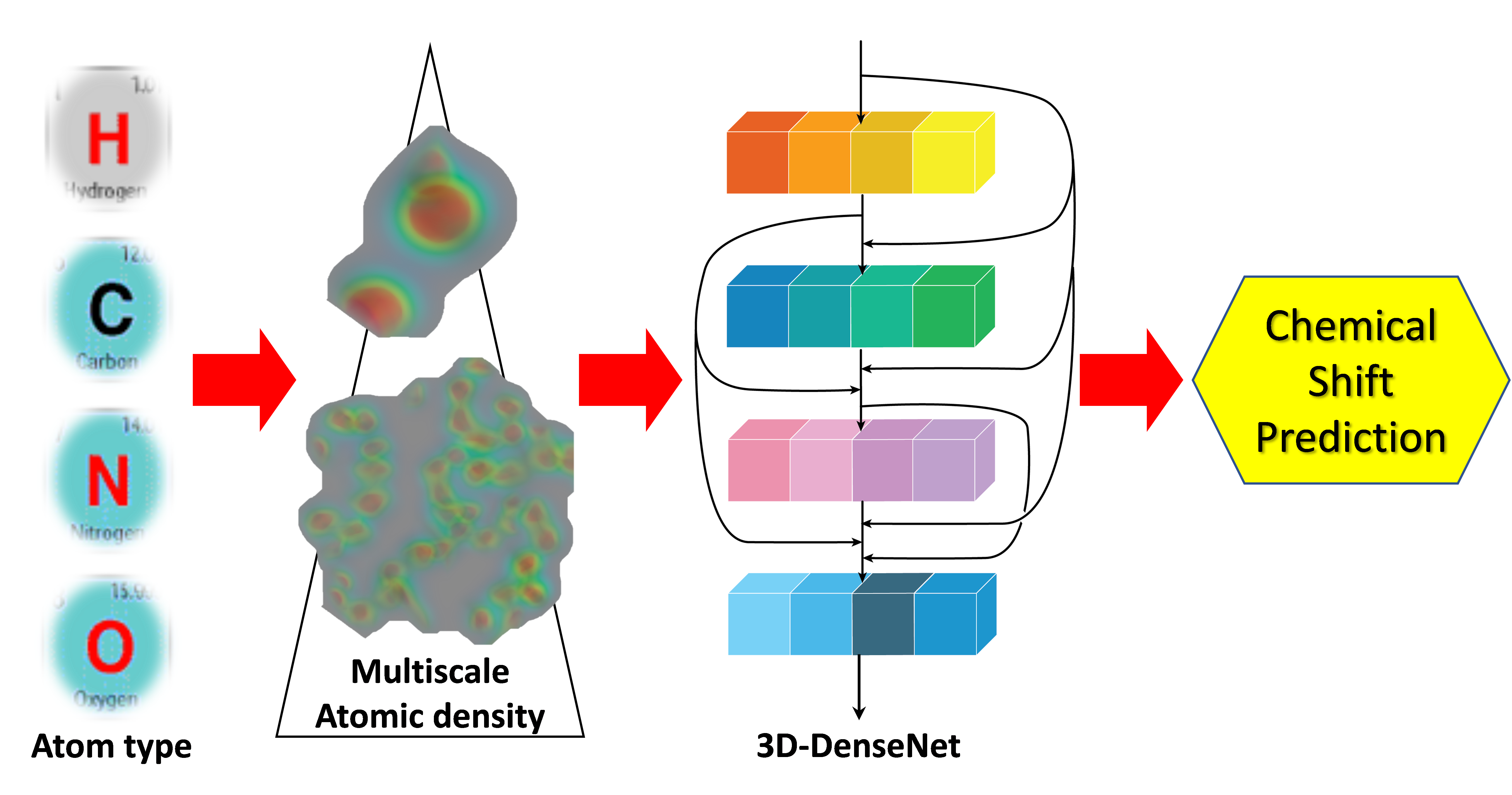}
\label{toc_graph}
\end{tocentry}

\begin{abstract}
We have developed a deep learning algorithm for chemical shift prediction for atoms in molecular crystals that utilizes an atom-centered Gaussian density model for the 3D data representation of a molecule. We define multiple channels that describe different spatial resolutions for each atom type that utilizes cropping, pooling, and concatenation to create a multi-resolution 3D-DenseNet architecture (MR-3D-DenseNet). Because the training and testing time scale linearly with the number of samples, the MR-3D-DenseNet can exploit data augmentation that takes into account the property of rotational invariance of the chemical shifts, thereby also increasing the size of the training dataset by an order of magnitude without additional cost. We obtain very good agreement for $^{13}$C, $^{15}$N, and $^{17}$O chemical shifts, with the highest accuracy found for $^1$H chemical shifts that is equivalent to the best predictions using \textit{ab initio} quantum chemistry methods.
\end{abstract}


\section{Introduction}
Nuclear magnetic resonance (NMR) crystallography is an experimental technique to determine the structure of complex materials \cite{nmr_mat1, nmr_mat2}, biomolecules such as proteins \cite{nmr_protein1, nmr_protein2}, as well as small molecules and pharmaceuticals \cite{nmr_pharm1, nmr_pharm2, nmr_pharm3} in the solid state. In practice, NMR crystallography is a structural model building procedure that depends on a number of NMR data types, of which chemical shifts in particular play a prominent role. A strength of NMR chemical shift data is its excellent sensitivity to hydrogen\cite{nmr_xrd} which, given the importance of hydrogen-bonding in most molecular systems, makes it very complementary to X-ray diffraction techniques. 

In the case where little is known about the chemical bonding of an unknown structure, the experimental measurements for chemical shifts are compared to the results of \textit{ab initio} methods based on density functional theory (DFT), typically using Gauge-Including Projector-Augmented Waves (GIPAW) methods \cite{gipaw}. However, because of the cubic computational complexity scaling with the number of atoms ($\mathcal{O}(N^3)$), alternative methods are being actively investigated to mitigate its large computational cost, especially for large systems. Many of these more inexpensive approaches are focused on fragment models that incorporate the long-range many-body polarization effects of the lattice environment via electrostatic embedding, such as the self-consistent reproduction of the Madelung potential (SCRMP)\cite{embed}. When combined with a DFT calculation within the cluster, this models has shown comparable results to GIPAW for chemical shift prediction.

An alternative approach is to apply machine learning methods to predict the experimental and/or DFT results for systems ranging from proteins in solution \cite{ml_prot1, ml_prot2, ml_prot3, ml_prot4, ml_prot5} to solid-state materials \cite{nmr_mat1, nmr_mat2, nmr_mat3}. Cuny et al.\ reported a fully connected shallow neural network to predict the quadrupolar couplings and chemical shifts in silica materials for $^{17}$O and $^{29}$Si using symmetric functions of the Cartesian coordinates as the input \cite{ml_nmr1}. Paruzzo et al. applied the kernel ridge regression (KRR) using a smooth overlap of atomic positions (SOAP) kernel, that also directly incorporates rotational invariance of the chemical shift value to applied magnetic field, for molecular crystal systems  \cite{ml_nmr2}.  However, the KRR approach requires $\mathcal{O}(N^2)$ complexity for calculating the similarity kernel matrix, and quadratic-to-cubic complexity for kernel matrix inversion, which is ultimately not tenable for large training and testing datasets.

Convolutional neural networks (CNNs) have been applied to several problems in chemistry and biology, such as enzyme classification\cite{cnn_enzyme}, molecular representation\cite{cnn_rep}, amino acid environment similarity\cite{cnn_aa}, and potential energy prediction\cite{cnn_energy}. They have not to the best of our knowledge been applied to NMR crystallography property prediction. There are a number of deep network variants that have been developed to address important deficiencies of a vanilla CNN, which are hard to train because of the vanishing (or exploding) gradient problem. This is because the repeated application of non-linear activation functions cause later outputs in the deep layers to flatten out, and back-propagated gradients are then diminished.

Residual networks (ResNets) were developed to precondition the network to learn the residual of a non-linear mapping by referencing it to an identity mapping, which is easier to train due to the presence in the network architecture of "identity shortcut connections".\cite{he2016} Because these network connections skip layers, there is more direct information flow from the loss function to correct the weight parameters of earlier layers. DenseNets build on these ideas by also utilizing skipped connections for better gradient flow, while at the same time also performing concatenation of feature maps that permits greater propagation and reuse of features in what is termed "deeper supervised learning".\cite{densenet}

Here we report a machine learning approach to predict chemical shifts in the solid-state for hydrogen ($^1$H), carbon ($^{13}$C), nitrogen ($^{15}$N) and oxygen ($^{17}$O) based on a multi-resolution (MR) spatial data representation, where each resolution level and atom type is formulated as an independent channel of a deep learning 3D-DenseNet architecture. We find that introducing concatenation of pooling layers (at reduced resolution) with cropping of feature maps (retaining high resolution features with reduced size) of the transformed data, combine with the MR input data representation to non-trivially contribute to the accuracy of chemical shift prediction. The resulting MR-3D-Densenet removes the restrictions imposed by KRR, i.e. the need to build in rotational invariance of chemical shifts as well as the limitations to small data sets\cite{ml_nmr2}, in order to take advantage of a data augmentation procedure in which we rotate the chemical environment for each atom in a sample, thereby increasing the data set size by close to an order of magnitude with little computational expense. 

Using the greater capacity of the MR-3D-DenseNet deep network, we obtain chemical shift prediction performance on all atom types that outperforms the previous KRR machine learning method. More importantly it now predicts $^1$H chemical shifts that are of the same quality as the best \textit{ab initio} predictions, with RMSE error of 0.37 ppm, while the chemical shift prediction of $^{13}$C, $^{15}$N, and $^{17}$O will undoubtedly improve further once more unique training samples are made available to exploit the deep network architecture. However, given the far better computational scaling of the multi-resolution 3D-DenseNet, we can afford to address this deficiency with much larger data sets than currently available in future studies.

\section{Results}
\subsection{Data Representation}
The molecular crystal structures are from the Cambridge Structural Database (CSD)\cite{CSD}, comprising 2,000 crystal structures in the training dataset and 500 crystal structures in the testing dataset. The coordinates of atoms in the unit cell, and the corresponding calculated chemical shieldings, are as given in the reported literature by Paruzzo and co-workers\cite{ml_nmr2}. This resulted in the number of unaugmented 3D samples for training and testing for each of the atom types as given in Table \ref{Table1}. No further data selection or cleaning procedures are applied to the original dataset, except that 0.05\% outliers (chemical shielding $<$ 0 or $>$ 40) in the $^1$H-NMR training dataset were removed.

\begin{table}
\begin{center}
\begin{tabular}{ c|c|c|c|c } 
 \hline\hline
 \multirow{3}{*}{Atom Type} & \multicolumn{4}{|c}{Number of Samples} \\  
 \cline{2-5}
 & \multicolumn{2}{|c|}{Training Dataset} & \multicolumn{2}{|c}{Testing Dataset} \\
 \cline{2-5}
 & w/o Augmentation & w/ Augmentation & w/o Augmentation & w/ Augmentation  \\
 \hline
 $^1$H & 76,174 & 609,392 & 29,913 & 239,304 \\
 $^{13}$C & 58,148 & 465,184 & 26,607 & 212,856 \\
 $^{15}$N & 27,814 & 222,512 & 2,713 & 21,704 \\
 $^{17}$O & 25,924 & 207,392 & 5,404 & 43,232 \\
 \hline\hline
\end{tabular}
\end{center}
\caption{The number of samples in training and testing datasets with and without data augmentation.}
\label{Table1}
\end{table}

Given the limited number of examples in the training dataset, we apply a physically motivated data augmentation method to improve the prediction performance of the MR-3D-DenseNet model. Since the chemical shift is invariant under rotational operations, we augment the data by rotating the Cartesian coordinates of atoms randomly with the Euler angles uniformly distributed between [$-\frac{\pi}{2}, \frac{\pi}{2}$] along each of $x$, $y$ and $z$ axis. During the training phase, both the original data and augmented data are included in the training dataset. During the testing phase, we average the prediction results among 8 different rotation configurations. The final number of training and testing examples after this augmentation are given in Table \ref{Table1}.

The input data representation to the MR-3D-DenseNet assumes that chemical shifts are sensitive to the electron density distribution of atoms in molecules. Hence a molecule is represented on a 3D grid in which each atom takes on a radial Gaussian density. The 3D image is a bounded box with $16 \times 16 \times 16$ voxels, with the density $D(\pmb{r})$ at each voxel taken as a sum of Gaussian distributions from all of the atoms
\begin{equation}
    D(\pmb{r})=\sum_{\pmb{r}'\in A}\exp(-\frac{||\pmb{r}-\pmb{r}'||^2}{\sigma^2})
\end{equation}
where the summation runs over atoms of a given atom type $A$ and the $\pmb{r}'$ are the corresponding atomic centers. The coordinate $\pmb{r}=(x,y,z)$ at the center of voxel (with index ($i,j,k$)) is calculated as 
\begin{equation}
    \pmb{r}=(x,y,z)=(\frac{(\frac{15}{2}-i)d}{15},\frac{(\frac{15}{2}-j)d}{15},\frac{(\frac{15}{2}-k)d}{15})
\end{equation}
where $d$ is the grid resolution. Unlike the Gaussian smearing method reported in literature \cite{cnn_rep}, we calculate the density at the center of the voxel numerically using 16-bit floating point numbers. 
We also considered additional electron density representations including Slater orbitals and calculated from the inverse Fourier transform of the atomic form factor, but found that they performed worse than the Gaussian representation that can be explained by their heavy tails (see supplementary information). 

\begin{figure}
\centering
\includegraphics[width=12cm]{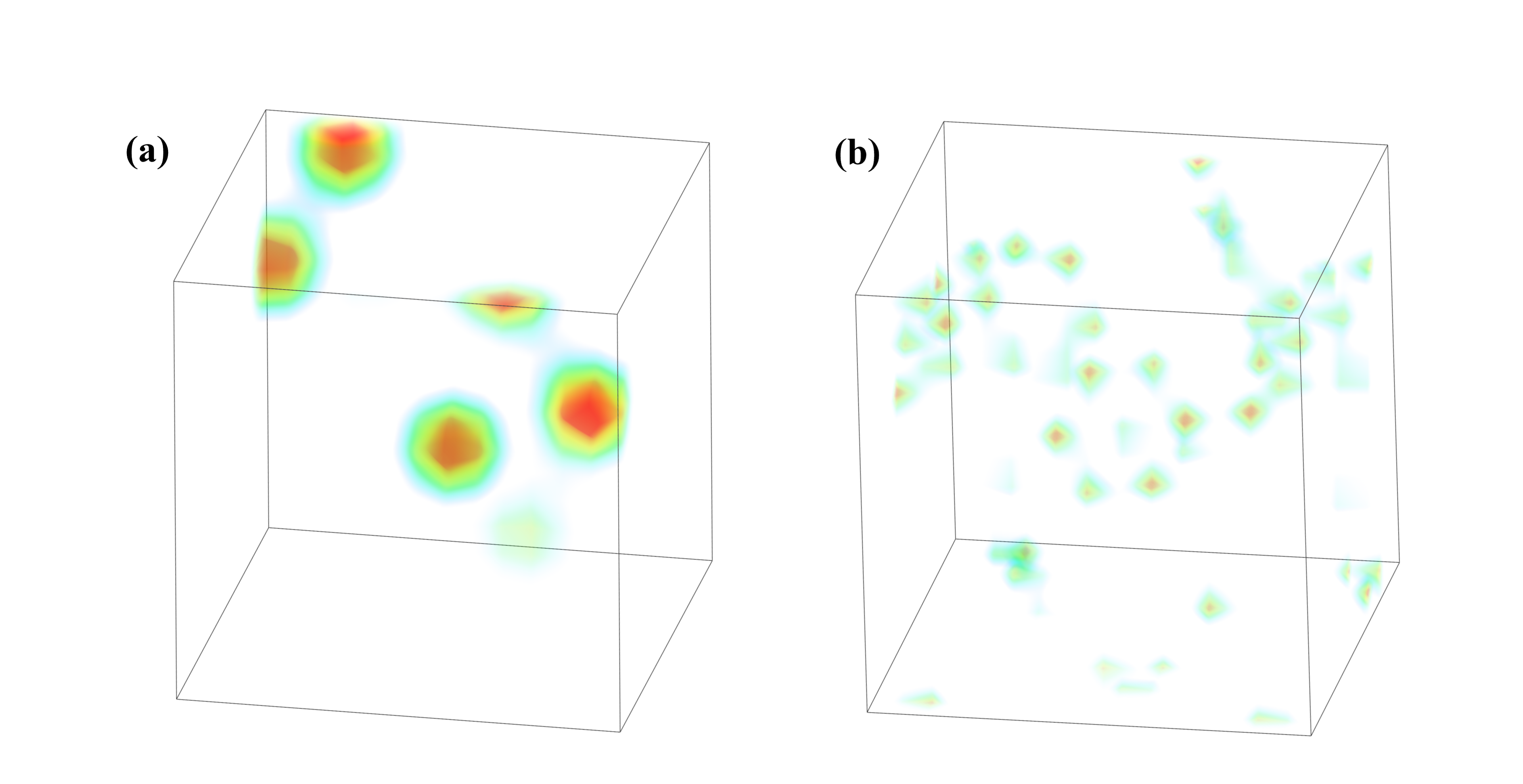}
\caption{Visualization of the Gaussian densities of atoms on different grid sizes. Representative example is shown for carbon channels on (a) 4 \AA{} and (b) 10 \AA{} grid. The densities are visualized through Mayavi package \cite{mayavi}.} 
\label{density_viz}
\end{figure}

The atom whose chemical shift is being evaluated is placed at the center of the 3D grid, and its chemical environment is represented by calculating the density under different grid sizes, where $d$ = 4 \AA, 6 \AA, 8 \AA, 10 \AA, and 14 \AA, each of which is represented by a dedicated channel in the MR-3D-DenseNet model. Under each grid size, we divide the density based on the atom types into 4 different channels for $^1$H, $^{13}$C, $^{15}$N, $^{17}$O, respectively, resulting in a total of 20 separate channels in the MR-3D-DenseNet network. Figure \ref{density_viz} shows a visualization of the carbon channels of the molecule (Z)-2-Hydroxy-3,3',4'-trimethoxystilbene (reported by Stomberg et al. \cite{cry_structure}) at two different grid size resolutions. 

\subsection{Machine Learning Models}
\begin{figure}
\centering
\includegraphics[width=\linewidth]{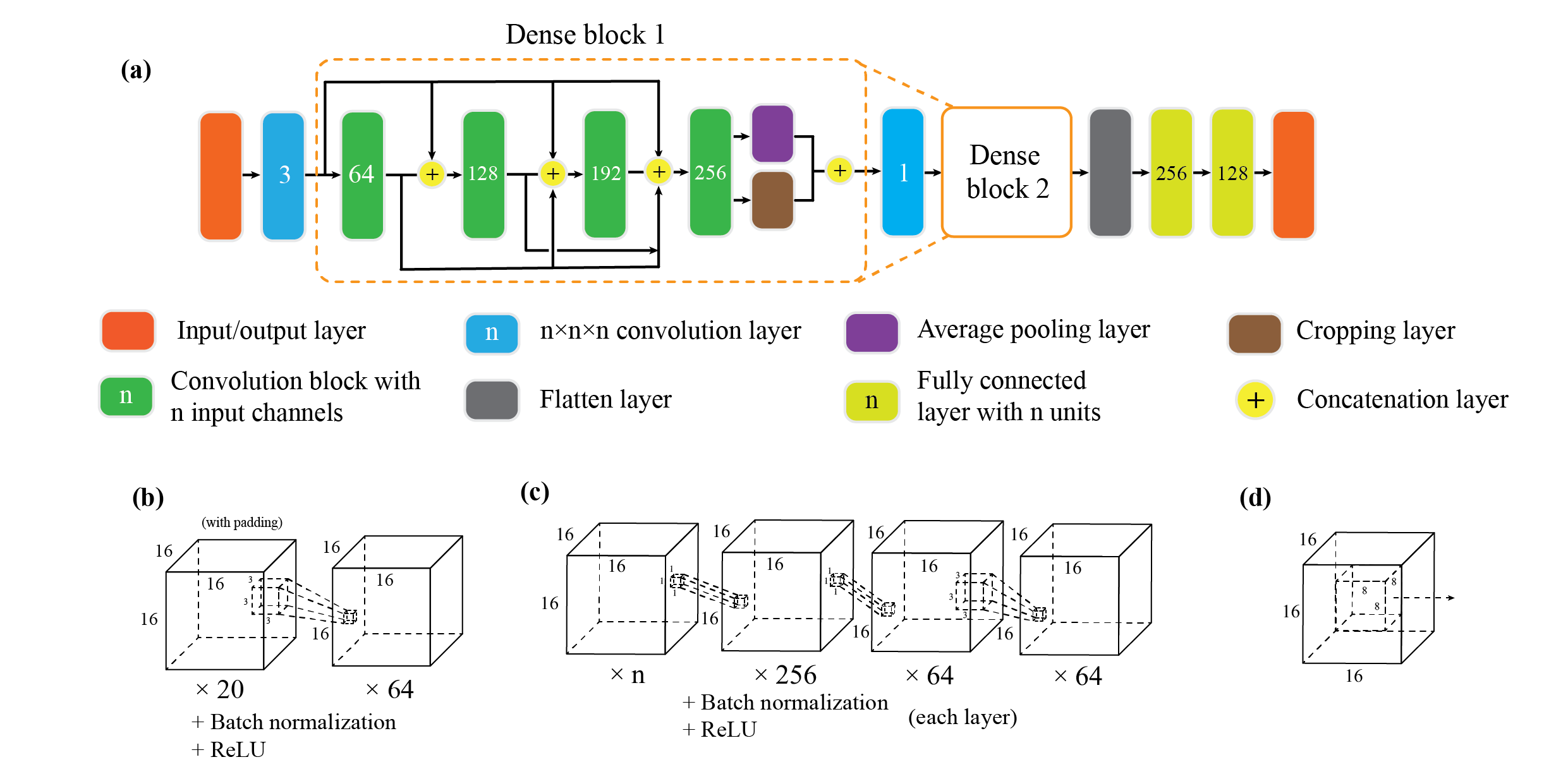}
\caption{Illustration of the overall architecture of the MR-3D-DenseNet model. (a) Flowchart of the network (b) Illustration of $3\times3\times3$ convolution layer prior to the first dense block (c) Illustration of the repeating unit in DenseNet block that contains two $1\times1\times1$ convolution layers followed by a $3\times3\times3$ convolution layer (d) Illustration of the cropping layer from the center of the feature map.}
\label{nn_archs}
\end{figure}

In this study, we designed a modification to a standard DenseNet that is motivated by the hypothesis that the importance of a given voxel increases as the distance between it and the investigated atom decreases, which is represented by multi-resolution channels. A schematic of the MR-3D-DenseNet architecture is shown in Figure \ref{nn_archs}, and is comprised of a regular $3\times3\times3$ convolutional layer followed by two DenseNet blocks with a $1\times1\times1$ transition convolutional layer in between them. The flattened output from the last DenseNet block is then fully connected to a layer with 256 units which is fully connected to a 128 unit layer, which is then fully connected to the output layer. Each DenseNet block has four repeating units: each repeating unit has two $1\times1\times1$ bottleneck convolutional layers with 256 and 64 channels followed by a $3\times3\times3$ convolution layer with 64 channels. The MR-3D-DenseNet utilizes cropping and pooling such that at the end of each block, we concatenate the $2\times2\times2$ average pooling layer and the cropping of the center segment of the feature map with the same size ($\frac{l}{2}$, where $l$ is the current feature map size). This retains low and high resolution features throughout the deep layers. Using this network architecture, we describe the detailed training protocol and hyperparameters in Methods. 

\subsection{Prediction Performance}
The performance on chemical shift predictions for all atoms using MR-3D-DenseNet compared to KRR is summarized in Table \ref{diff_atom}. The testing RMSEs of chemical shifts for $^1$H, $^{13}$C, $^{15}$N, and $^{17}$O using the MR-3D-DenseNet architecture is found to be 0.37 ppm, 3.3 ppm, 10.2 ppm and 15.3 ppm, which are $24\%, 23\%, 23\%$ and $14\%$ lower than the RMSEs given by a KRR method\cite{ml_nmr2}. Among the four atom types, the prediction performance of $^1$H is competitive in quality to both full and embedded \textit{ab initio} methods, with GIPAW/PBE and SCRMP/PBE0 having RMSEs of 0.43 ppm and 0.33 ppm respectively.\cite{embed, gipaw_pbe} Although the predictions on the other atom types are very good, we attribute their lessened performance with respect to \textit{ab initio} models as a lack of unique data compared to that available for $^1$H (Table \ref{Table1}), a point to which we return to later.

\begin{table}
\begin{center}
\begin{tabular}{ c|c|c } 
 \hline\hline
 Atom Type &  MR-3D-DenseNet & $R^2$ \\
 \hline
 H & 0.37 (24\%) & 0.9856 \\ 
 C & 3.3 (23\%) & 0.9957 \\
 N & 10.2 (23\%) & 0.9916 \\
 O & 15.3 (14\%) & 0.9933 \\
 \hline\hline
\end{tabular}
\end{center}
\caption{Testing RMSEs (ppm) using MR-3D-DenseNet. We also report the improvement of RMSE in percentage compared to KRR\cite{ml_nmr2} and the $R^2$ values using MR-3D-DenseNet.}
\label{diff_atom}
\end{table}

In a separate publication, we will present a full study of different deep learning architectures, but here we contrast the best MR-3D-DenseNet model to the KRR machine learning method for which results are available on the same chemical shift problem\cite{ml_nmr2}. We can attribute the success of the MR-3D-DenseNet approach based on three factors: (1) the greater flexibility in input representation of individual atom types and spatial resolution, and the advantages of concatenation of the pooling and cropping operations in the architecture, (2) the dependence on the size and quality of the training set, and (3) the ability to learn chemical bonding features, all of which are unique to chemical shift prediction using the MR-3D-DenseNet architecture.

In regards the first point we decompose the MR-3D-DenseNet result based on its multi-resolution input representation (MR-Input) with no special concatenation of pooling and cropping operations vs. the multi-resolution architecture (MR-Arch) that utilizes pooling and cropping but takes in only a single resolution input representation. It is evident that the input and architecture features trained in isolation of each other offer significant improvements in performance over KRR, with further benefit being realized by their combined used in MR-3D-DenseNet (Table \ref{multi_res}). The main limitation of the MR-Arch model is that the 3D-grid size of the single resolution input depends on the atom under consideration. In contrast, using a MR-Input model allows different resolution grid sizes to be easily combined without pre-screening for each atom type. Moreover, combining MR-Input with the concatenation of the pooling at lower resolutions with the feature map close to the investigated atom at higher resolution, which provides a pathway for information to flow through the intermediate feature space for the full MR-3D-DenseNet, to arrive at the best prediction performance. 

\begin{table}
\begin{center}
\begin{tabular}{ c|c|c|c|c } 
 \hline\hline
 Atom Type & KRR & MR-Arch & MR-Input & 3D-MR-DenseNet \\
 \hline
 H & 0.49 & 0.38 (10 \AA) & 0.38 & 0.37 \\
 C & 4.3 & 3.5 (6 \AA) & 3.5 & 3.3 \\
 N & 13.3 & 10.2 (8 \AA) & 10.3 & 10.2 \\
 O & 17.7 & 16.3 (6 \AA) & 15.6 & 15.3 \\
 \hline\hline
\end{tabular}
\end{center}
\caption{Testing RMSEs (ppm) for KRR and using different multi-resolution features of the MR-3D-DenseNet model for each atom type: MR-Arch, MR-Input, and MR-3D-DenseNet. For the single-resolution input, the MR-Arch model is sensitive to the grid size for a given atom type (parentheses).}
\label{multi_res}
\end{table}

\begin{figure}
\centering
\includegraphics[width=14cm]{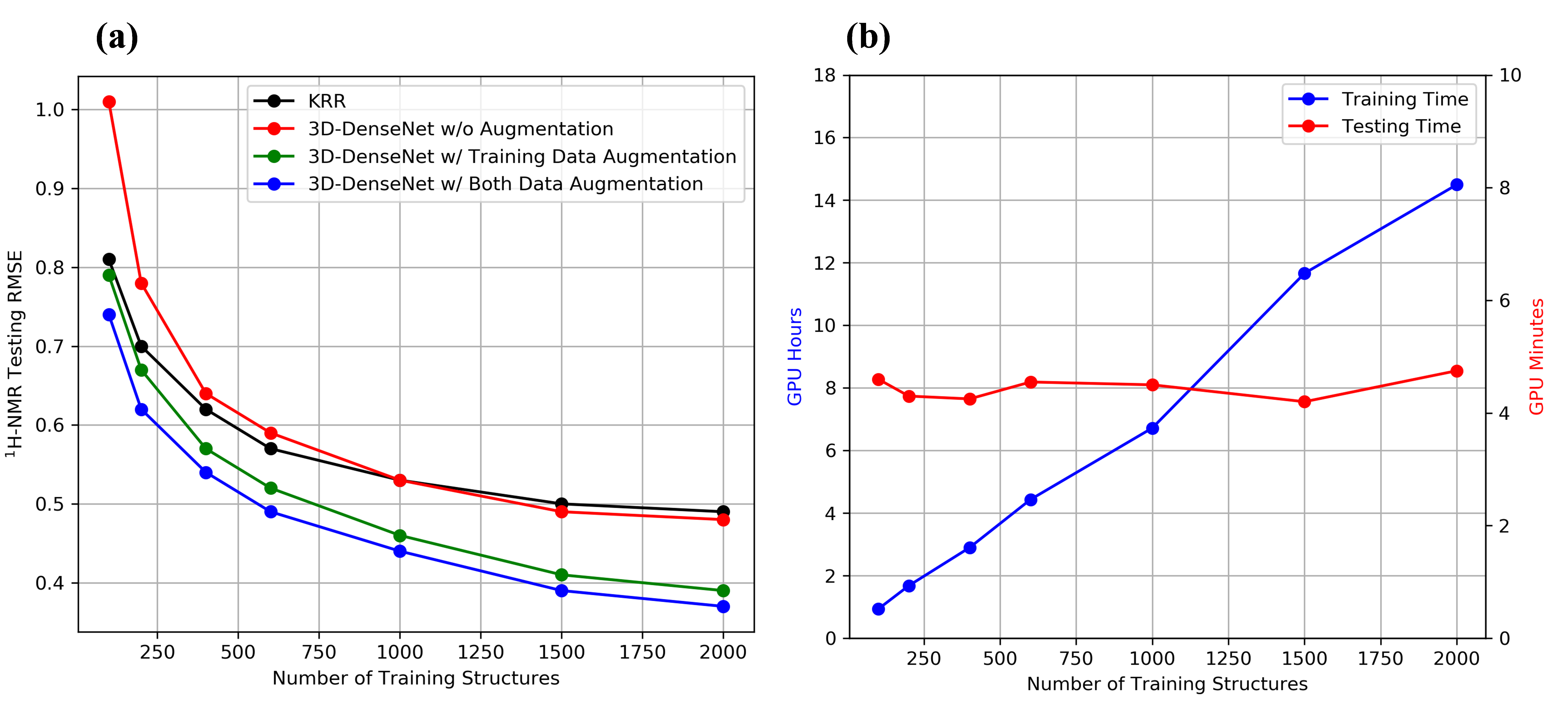}
\caption{Testing RMSEs and timings for $^1$H chemical shift for different numbers of samples using the MR-3D-DenseNet. (a) using no augmentation (red), with training dataset 8-fold augmentation (green), using both training and testing dataset with 8-fold augmentation (blue), and compared to the testing error reported previously for KRR on the same dataset\cite{ml_nmr2} (black). The models are trained under the same number of batches to obtain a fair comparison; for example, when the data is augmented by 8-fold, the number of training epochs and the learning rate decay decrease to 1/8.  (b) Training (8-fold) time of MR-3D-DenseNet model for the $^1$H chemical shift. The testing time (1-fold) of $^1$H chemical shift is about 4-5 minutes for 500 testing structures and is independent on the number of training structures. The training and testing time are benchmarked on Nvidia Tesla P100 GPU.}
\label{pred_scale}
\end{figure}

Furthermore, the success of any neural network model is highly dependent on the size, variety and quality of the training dataset. To understand the effect of the training data size, we examine the $^1$H chemical shift testing RMSE for KRR and the MR-3D-DenseNet model as a function of increasing number of training examples (Figure \ref{pred_scale}a). As expected the prediction performance of MR-3D-DenseNet ultimately improves over KRR when 1000's of samples are added into the training dataset. However, the  MR-3D-DenseNet has the capacity to exploit the augmented data to outperform the KRR model even with only 100 training structures. Although it might be argued that the KRR model has no need for the augmented data, since rotational invariance is built directly into the kernel, the data augmentation is clearly doing something more than invoking the rotational invariance feature of the chemical shift (i.e. the performance would be the same otherwise). In addition, augmentation of the testing dataset can be seen as equivalent to an ensemble averaging prediction without the need to retrain many networks to realize the same benefit, lowering the testing RMSE  further to realize the best MR-3D-DenseNet performance (Figure \ref{pred_scale}a).

\begin{figure}
\centering
\includegraphics[width=14cm]{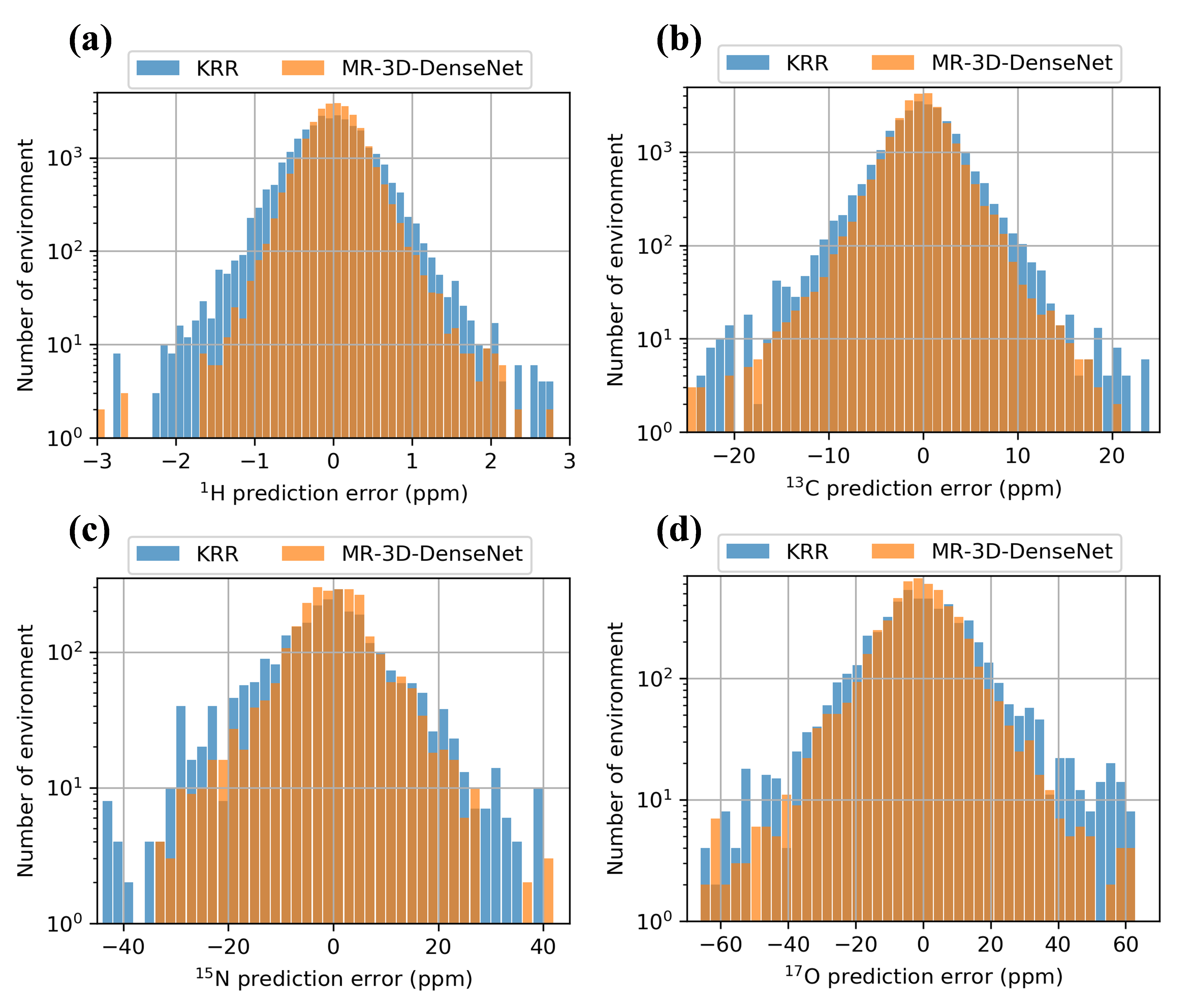}
\caption{Histogram of testing error distribution comparing MR-3D-DenseNet and KRR for (a) $^1$H, (b) $^{13}$C, (c) $^{15}$N and (d) $^{17}$O} 
\label{pred_dist}
\end{figure}

How to discover what the data augmentation is providing can't be easily captured with KRR due its unfavorable computational scaling for kernel matrix computation and kernel matrix inversion. By contrast the training time for the MR-3D-DenseNet model scales linearly with the number of training samples (Figure \ref{pred_scale}b). More importantly, the prediction time of MR-3D-DenseNet with a trained model does not scale with the number of training examples, whereas the testing time for KRR scales linearly because the similarity kernel has to be calculated using all of the training samples. In totality, the MR-3D-DenseNet architecture with data augmentation yields a much tighter prediction error across the unique data across all atom types relative to KRR as seen in Figure \ref{pred_dist}. We found that further increasing the data augmentation to 16-fold rotations or adding the effects of small vibrational smearing of atom positions had a neutral effect on the prediction performance. Instead Figure \ref{pred_dist} emphasizes that creating more unique data for the heavy atoms will certainly improve the MR-3D-DenseNet performance relative to \textit{ab initio} models, as the number of heavy atom samples are limited compared to $^1$H samples in the current dataset. 

\begin{figure}
\centering
\includegraphics[width=14cm]{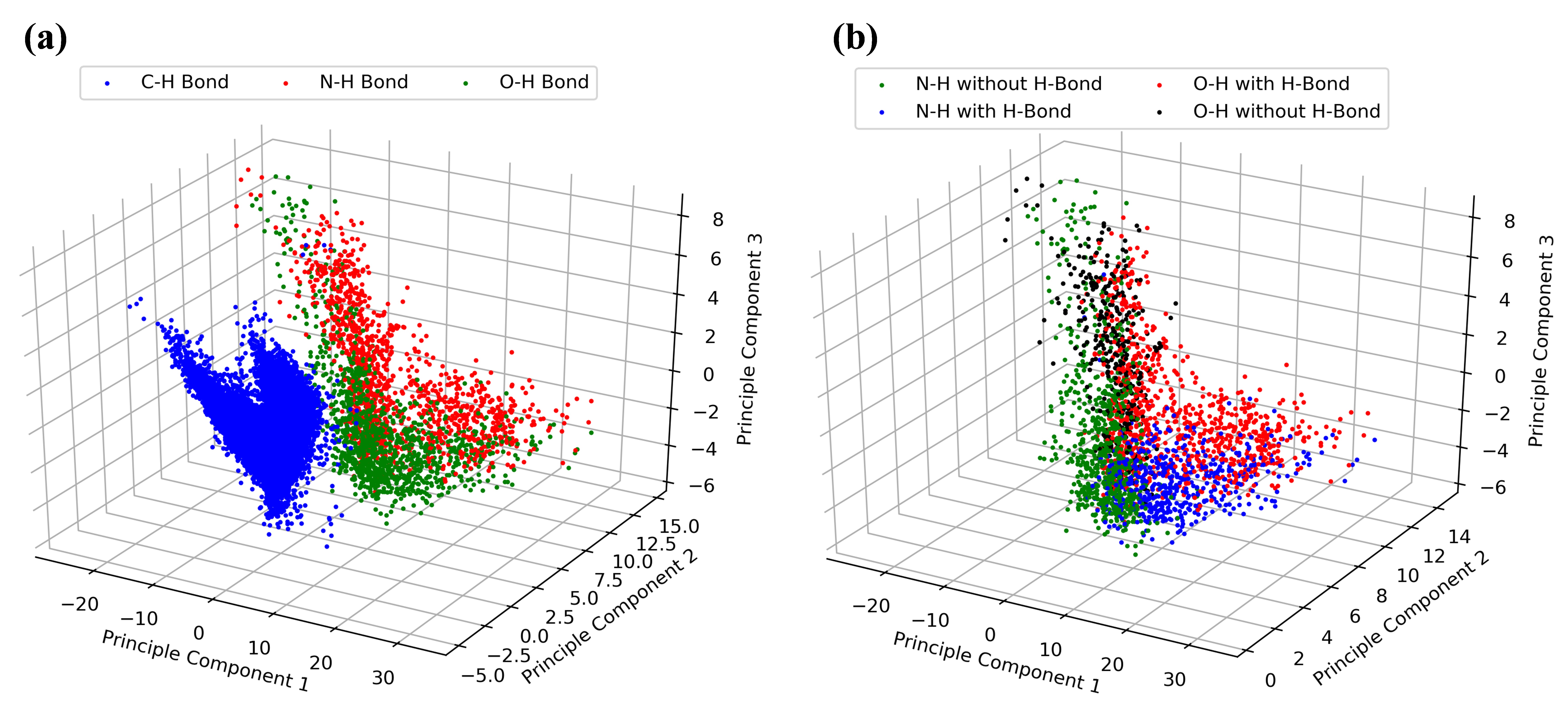}
\caption{Visualization of the data in the last fully connected layer by projecting the data into 3D space using principle component analysis (PCA). It shows the clustering of different (a) chemical bonds and (b) hydrogen bonds.}
\label{pca}
\end{figure}

The MR-3D-DenseNet model is interpretable by extracting the chemical bonding and hydrogen-bonding information that is clearly relevant for $^1$H chemical shift prediction using principle component analysis (PCA). We visualize the distribution of the data in the last fully connected layer by applying PCA to project the data onto a 3D space as shown in Figure \ref{pca}. The first three components can explain $>95\%$ of the variances of the data (using the explained ratio with different number of component as plotted in Figure S2 in the supplementary information). Even though no explicit bonding information was provided as the input of the neural network, the model is capable of separating the C-H, N-H, and O-H bond clusters. Furthermore, for N-H and O-H bonds, the hydrogen which or do not act as a hydrogen-bonding donor are clustered individually. This analysis proves that the 3D-DenseNet can extract relevant chemical interpretations for chemical shifts, similar to what has been reported for other structural properties in proteins\cite{ml_prot1, ml_prot2, ml_prot3, ml_prot4, ml_prot5}.

\section{Conclusion}
We have presented a 3D-DenseNet deep learning model that exploits a multi-resolution approach for NMR chemical shift prediction of atoms in molecular crystals. A unique feature of our deep learning model is the use of multi-resolution spatial input data organized into individualized channels for each atom type, that provides a learning framework for different tasks which are sensitive to the density representation with different cut-off sizes. Furthermore, the multi-resolution architecture combines the benefits of both pooling and the high-resolution feature map close to the investigated atom, which can potentially be applied to the prediction of other chemical properties sensitive to different length scales. 

In addition to its greater flexibility in representing spatial density distributions, the 3D-DenseNet can more efficiently handle much larger data sets. In comparison to the KRR approach, the 3D-DenseNet method has the capacity and favorable scaling characteristics that allowed us to increase the training data by an order of magnitude through rotation of the input samples to predict chemical shift values based in part on the rotational invariance property of chemical shifts. As a result, the totality of our deep learning approach can predict the chemical shifts more accurately, especially for $^1$H chemical shifts which has a testing error that is equivalent to the highest level of chemical accuracy possible with \textit{ab initio} methods.

The accurate chemical shift prediction of $^1$H is important for the structure characterization of many solid-state chemistry and biological systems as NMR crystallography is one of the most powerful techniques to study the structure and dynamics of hydrogen atoms in solid-state under natural abundance. Although our chemical shift prediction for other heavy atom data types were significantly improved with respect to KRR, it does not reach the same level of accuracy, almost certainly due to the more limited amount of data available that prevents us exploiting the capacity of the MR-3D-DenseNet. This highlights the importance of the size and diversity of training dataset, which will be a topic of future work for chemical shift prediction for $^{13}$C, $^{15}$N and $^{17}$O.   

\section{Methods}
The neural network is implemented using Keras\cite{chollet2015keras} with Tensorflow\cite{tensorflow2015-whitepaper} as the backend. The neural network architecture and hyperparameters are optimized through a 4-fold cross-validation on the training dataset. We trained four dedicated models for the chemical shift prediction of $^1$H, $^{13}$C, $^{15}$N and $^{17}$O-NMR separately. To accelerate the convergence, we subtract the mean of the chemical shielding values and divide them by $1, 10, 30, 40$ for $^1$H, $^{13}$C, $^{15}$N and $^{17}$O-NMR respectively during the training phase. The mean and scaling factors are applied back during the testing phase.  Each layer is followed by a batch normalization (BatchNorm\cite{BatchNorm}) layer, and a rectified linear units (ReLU\cite{RELU}) layer. There are two dropout layers added after each fully connected layers with rate 0.1. The $L_2$ regularizer with $\lambda=3\times 10^{-5}$ is applied to all the weights in the neural network. The training epochs used are 12, 15, 24 and 24, and the decay rates $\alpha$ are 0.6, 0.5, 0.25 and 0.25 for $^1$H, $^{13}$C, $^{15}$N and $^{17}$O, respectively. The batch size is fixed to 128. The learning rate starts with $10^{-3}$ and decays exponentially. In epoch $i$, the learning rate is decayed to $10^{-3}\times\exp(-i\alpha)$. The hyperparameter details are summarized in Table \ref{hyper_params}. The testing RMSEs are reported by averaging the results from at least three experiments in which the models are initialized with different random seed and the training data are shuffled in random order in each training epoch. The standard deviations of RMSEs are 0.0026 ppm, 0.038 ppm, 0.08 ppm and 0.27 ppm.

\begin{table}
\begin{center}
\begin{tabular}{ c|c|c } 
 \hline\hline 
 Atom Type & Number of Training Epoch & Decay Rate $\alpha$\\
 \hline
 $^1$H & 12 & 0.6\\
 $^{13}$C & 15 & 0.5\\
 $^{15}$N & 24 & 0.25\\
 $^{17}$O & 24 & 0.25\\
 \hline\hline
\end{tabular}
\end{center}
\caption{Number of epochs and learning rate decay.}
\label{hyper_params}
\end{table}

\section{Acknowledgments} We thank the National Institutes of Health for support under Grant No. 5U01GM121667. KCB thanks the California Alliance Postdoctoral Fellowship for early support of his work on this project. This research used the computational resources of the National Energy Research Scientific Computing Center, a DOE Office of Science User Facility supported by the Office of Science of the U.S. Department of Energy under Contract No. DE-AC02-05CH11231.

\bibliographystyle{unsrt}
\bibliography{nmr}

\providecommand{\latin}[1]{#1}
\makeatletter
\providecommand{\doi}
  {\begingroup\let\do\@makeother\dospecials
  \catcode`\{=1 \catcode`\}=2 \doi@aux}
\providecommand{\doi@aux}[1]{\endgroup\texttt{#1}}
\makeatother
\providecommand*\mcitethebibliography{\thebibliography}
\csname @ifundefined\endcsname{endmcitethebibliography}
  {\let\endmcitethebibliography\endthebibliography}{}
\begin{mcitethebibliography}{33}
\providecommand*\natexlab[1]{#1}
\providecommand*\mciteSetBstSublistMode[1]{}
\providecommand*\mciteSetBstMaxWidthForm[2]{}
\providecommand*\mciteBstWouldAddEndPuncttrue
  {\def\EndOfBibitem{\unskip.}}
\providecommand*\mciteBstWouldAddEndPunctfalse
  {\let\EndOfBibitem\relax}
\providecommand*\mciteSetBstMidEndSepPunct[3]{}
\providecommand*\mciteSetBstSublistLabelBeginEnd[3]{}
\providecommand*\EndOfBibitem{}
\mciteSetBstSublistMode{f}
\mciteSetBstMaxWidthForm{subitem}{(\alph{mcitesubitemcount})}
\mciteSetBstSublistLabelBeginEnd
  {\mcitemaxwidthsubitemform\space}
  {\relax}
  {\relax}

\bibitem[Martineau(2014)]{nmr_mat1}
Martineau,~C. NMR crystallography: Applications to inorganic materials.
  \emph{Solid State Nuclear Magnetic Resonance} \textbf{2014}, \emph{63-64}, 1
  -- 12\relax
\mciteBstWouldAddEndPuncttrue
\mciteSetBstMidEndSepPunct{\mcitedefaultmidpunct}
{\mcitedefaultendpunct}{\mcitedefaultseppunct}\relax
\EndOfBibitem
\bibitem[Bryce(2017)]{nmr_mat2}
Bryce,~D.~L. NMR crystallography: structure and properties of materials from
  solid-state nuclear magnetic resonance observables. \emph{IUCrJ}
  \textbf{2017}, \emph{4}, 350--359\relax
\mciteBstWouldAddEndPuncttrue
\mciteSetBstMidEndSepPunct{\mcitedefaultmidpunct}
{\mcitedefaultendpunct}{\mcitedefaultseppunct}\relax
\EndOfBibitem
\bibitem[Shahid \latin{et~al.}(2012)Shahid, Bardiaux, Franks, Krabben, Habeck,
  van Rossum, and Linke]{nmr_protein1}
Shahid,~S.~A.; Bardiaux,~B.; Franks,~W.~T.; Krabben,~L.; Habeck,~M.; van
  Rossum,~B.-J.; Linke,~D. Membrane-protein structure determination by
  solid-state NMR spectroscopy of microcrystals. \emph{Nature methods}
  \textbf{2012}, \emph{9}, 1212\relax
\mciteBstWouldAddEndPuncttrue
\mciteSetBstMidEndSepPunct{\mcitedefaultmidpunct}
{\mcitedefaultendpunct}{\mcitedefaultseppunct}\relax
\EndOfBibitem
\bibitem[Macholl \latin{et~al.}(2013)Macholl, Tietze, and
  Buntkowsky]{nmr_protein2}
Macholl,~S.; Tietze,~D.; Buntkowsky,~G. NMR crystallography of amides, peptides
  and protein--ligand complexes. \emph{CrystEngComm} \textbf{2013}, \emph{15},
  8627--8638\relax
\mciteBstWouldAddEndPuncttrue
\mciteSetBstMidEndSepPunct{\mcitedefaultmidpunct}
{\mcitedefaultendpunct}{\mcitedefaultseppunct}\relax
\EndOfBibitem
\bibitem[Abraham \latin{et~al.}(2016)Abraham, Apperley, Byard, Ilott, Robbins,
  Zorin, Harris, and Hodgkinson]{nmr_pharm1}
Abraham,~A.; Apperley,~D.~C.; Byard,~S.~J.; Ilott,~A.~J.; Robbins,~A.~J.;
  Zorin,~V.; Harris,~R.~K.; Hodgkinson,~P. Characterising the role of water in
  sildenafil citrate by NMR crystallography. \emph{CrystEngComm} \textbf{2016},
  \emph{18}, 1054--1063\relax
\mciteBstWouldAddEndPuncttrue
\mciteSetBstMidEndSepPunct{\mcitedefaultmidpunct}
{\mcitedefaultendpunct}{\mcitedefaultseppunct}\relax
\EndOfBibitem
\bibitem[Skotnicki \latin{et~al.}(2015)Skotnicki, Apperley, Aguilar,
  Milanowski, Pyda, and Hodgkinson]{nmr_pharm2}
Skotnicki,~M.; Apperley,~D.~C.; Aguilar,~J.~A.; Milanowski,~B.; Pyda,~M.;
  Hodgkinson,~P. Characterization of two distinct amorphous forms of valsartan
  by solid-state NMR. \emph{Molecular pharmaceutics} \textbf{2015}, \emph{13},
  211--222\relax
\mciteBstWouldAddEndPuncttrue
\mciteSetBstMidEndSepPunct{\mcitedefaultmidpunct}
{\mcitedefaultendpunct}{\mcitedefaultseppunct}\relax
\EndOfBibitem
\bibitem[Baias \latin{et~al.}(2013)Baias, Widdifield, Dumez, Thompson, Cooper,
  Salager, Bassil, Stein, Lesage, Day, \latin{et~al.} others]{nmr_pharm3}
Baias,~M.; Widdifield,~C.~M.; Dumez,~J.-N.; Thompson,~H.~P.; Cooper,~T.~G.;
  Salager,~E.; Bassil,~S.; Stein,~R.~S.; Lesage,~A.; Day,~G.~M. \latin{et~al.}
  Powder crystallography of pharmaceutical materials by combined crystal
  structure prediction and solid-state 1H NMR spectroscopy. \emph{Physical
  Chemistry Chemical Physics} \textbf{2013}, \emph{15}, 8069--8080\relax
\mciteBstWouldAddEndPuncttrue
\mciteSetBstMidEndSepPunct{\mcitedefaultmidpunct}
{\mcitedefaultendpunct}{\mcitedefaultseppunct}\relax
\EndOfBibitem
\bibitem[Cui \latin{et~al.}(2019)Cui, Olmsted, Mehta, Asta, and Hayes]{nmr_xrd}
Cui,~J.; Olmsted,~D.~L.; Mehta,~A.~K.; Asta,~M.; Hayes,~S.~E. NMR
  Crystallography: Evaluation of Hydrogen Positions in Hydromagnesite by 13C
  $\{$1H$\}$ REDOR Solid-State NMR and Density Functional Theory Calculation of
  Chemical Shielding Tensors. \emph{Angewandte Chemie International Edition}
  \textbf{2019}, \emph{58}, 4210--4216\relax
\mciteBstWouldAddEndPuncttrue
\mciteSetBstMidEndSepPunct{\mcitedefaultmidpunct}
{\mcitedefaultendpunct}{\mcitedefaultseppunct}\relax
\EndOfBibitem
\bibitem[Pickard and Mauri(2001)Pickard, and Mauri]{gipaw}
Pickard,~C.~J.; Mauri,~F. All-electron magnetic response with pseudopotentials:
  NMR chemical shifts. \emph{Physical Review B} \textbf{2001}, \emph{63},
  245101\relax
\mciteBstWouldAddEndPuncttrue
\mciteSetBstMidEndSepPunct{\mcitedefaultmidpunct}
{\mcitedefaultendpunct}{\mcitedefaultseppunct}\relax
\EndOfBibitem
\bibitem[Hartman \latin{et~al.}(2017)Hartman, Balaji, and Beran]{embed}
Hartman,~J.~D.; Balaji,~A.; Beran,~G. J.~O. Improved Electrostatic Embedding
  for Fragment-Based Chemical Shift Calculations in Molecular Crystals.
  \emph{Journal of Chemical Theory and Computation} \textbf{2017}, \emph{13},
  6043--6051, PMID: 29139294\relax
\mciteBstWouldAddEndPuncttrue
\mciteSetBstMidEndSepPunct{\mcitedefaultmidpunct}
{\mcitedefaultendpunct}{\mcitedefaultseppunct}\relax
\EndOfBibitem
\bibitem[Shen and Bax(2007)Shen, and Bax]{ml_prot1}
Shen,~Y.; Bax,~A. Protein backbone chemical shifts predicted from searching a
  database for torsion angle and sequence homology. \emph{Journal of
  biomolecular NMR} \textbf{2007}, \emph{38}, 289--302\relax
\mciteBstWouldAddEndPuncttrue
\mciteSetBstMidEndSepPunct{\mcitedefaultmidpunct}
{\mcitedefaultendpunct}{\mcitedefaultseppunct}\relax
\EndOfBibitem
\bibitem[Shen and Bax(2010)Shen, and Bax]{ml_prot2}
Shen,~Y.; Bax,~A. SPARTA+: a modest improvement in empirical NMR chemical shift
  prediction by means of an artificial neural network. \emph{Journal of
  biomolecular NMR} \textbf{2010}, \emph{48}, 13--22\relax
\mciteBstWouldAddEndPuncttrue
\mciteSetBstMidEndSepPunct{\mcitedefaultmidpunct}
{\mcitedefaultendpunct}{\mcitedefaultseppunct}\relax
\EndOfBibitem
\bibitem[Neal \latin{et~al.}(2003)Neal, Nip, Zhang, and Wishart]{ml_prot3}
Neal,~S.; Nip,~A.~M.; Zhang,~H.; Wishart,~D.~S. Rapid and accurate calculation
  of protein 1 H, 13 C and 15 N chemical shifts. \emph{Journal of biomolecular
  NMR} \textbf{2003}, \emph{26}, 215--240\relax
\mciteBstWouldAddEndPuncttrue
\mciteSetBstMidEndSepPunct{\mcitedefaultmidpunct}
{\mcitedefaultendpunct}{\mcitedefaultseppunct}\relax
\EndOfBibitem
\bibitem[Han \latin{et~al.}(2011)Han, Liu, Ginzinger, and Wishart]{ml_prot4}
Han,~B.; Liu,~Y.; Ginzinger,~S.~W.; Wishart,~D.~S. SHIFTX2: significantly
  improved protein chemical shift prediction. \emph{Journal of biomolecular
  NMR} \textbf{2011}, \emph{50}, 43\relax
\mciteBstWouldAddEndPuncttrue
\mciteSetBstMidEndSepPunct{\mcitedefaultmidpunct}
{\mcitedefaultendpunct}{\mcitedefaultseppunct}\relax
\EndOfBibitem
\bibitem[Kohlhoff \latin{et~al.}(2009)Kohlhoff, Robustelli, Cavalli,
  Salvatella, and Vendruscolo]{ml_prot5}
Kohlhoff,~K.~J.; Robustelli,~P.; Cavalli,~A.; Salvatella,~X.; Vendruscolo,~M.
  Fast and accurate predictions of protein NMR chemical shifts from interatomic
  distances. \emph{Journal of the American Chemical Society} \textbf{2009},
  \emph{131}, 13894--13895\relax
\mciteBstWouldAddEndPuncttrue
\mciteSetBstMidEndSepPunct{\mcitedefaultmidpunct}
{\mcitedefaultendpunct}{\mcitedefaultseppunct}\relax
\EndOfBibitem
\bibitem[Leclaire \latin{et~al.}(2016)Leclaire, Poisson, Ziarelli, Pepe,
  Fotiadu, Paruzzo, Rossini, Dumez, Elena-Herrmann, and Emsley]{nmr_mat3}
Leclaire,~J.; Poisson,~G.; Ziarelli,~F.; Pepe,~G.; Fotiadu,~F.; Paruzzo,~F.~M.;
  Rossini,~A.~J.; Dumez,~J.-N.; Elena-Herrmann,~B.; Emsley,~L. Structure
  elucidation of a complex CO 2-based organic framework material by NMR
  crystallography. \emph{Chemical science} \textbf{2016}, \emph{7},
  4379--4390\relax
\mciteBstWouldAddEndPuncttrue
\mciteSetBstMidEndSepPunct{\mcitedefaultmidpunct}
{\mcitedefaultendpunct}{\mcitedefaultseppunct}\relax
\EndOfBibitem
\bibitem[Cuny \latin{et~al.}(2016)Cuny, Xie, Pickard, and Hassanali]{ml_nmr1}
Cuny,~J.; Xie,~Y.; Pickard,~C.~J.; Hassanali,~A.~A. Ab initio quality NMR
  parameters in solid-state materials using a high-dimensional neural-network
  representation. \emph{Journal of chemical theory and computation}
  \textbf{2016}, \emph{12}, 765--773\relax
\mciteBstWouldAddEndPuncttrue
\mciteSetBstMidEndSepPunct{\mcitedefaultmidpunct}
{\mcitedefaultendpunct}{\mcitedefaultseppunct}\relax
\EndOfBibitem
\bibitem[Paruzzo \latin{et~al.}(2018)Paruzzo, Hofstetter, Musil, De, Ceriotti,
  and Emsley]{ml_nmr2}
Paruzzo,~F.~M.; Hofstetter,~A.; Musil,~F.; De,~S.; Ceriotti,~M.; Emsley,~L.
  Chemical shifts in molecular solids by machine learning. \emph{Nature
  communications} \textbf{2018}, \emph{9}, 4501\relax
\mciteBstWouldAddEndPuncttrue
\mciteSetBstMidEndSepPunct{\mcitedefaultmidpunct}
{\mcitedefaultendpunct}{\mcitedefaultseppunct}\relax
\EndOfBibitem
\bibitem[Amidi \latin{et~al.}(2018)Amidi, Amidi, Vlachakis, Megalooikonomou,
  Paragios, and Zacharaki]{cnn_enzyme}
Amidi,~A.; Amidi,~S.; Vlachakis,~D.; Megalooikonomou,~V.; Paragios,~N.;
  Zacharaki,~E.~I. EnzyNet: enzyme classification using 3D convolutional neural
  networks on spatial representation. \emph{PeerJ} \textbf{2018}, \emph{6},
  e4750\relax
\mciteBstWouldAddEndPuncttrue
\mciteSetBstMidEndSepPunct{\mcitedefaultmidpunct}
{\mcitedefaultendpunct}{\mcitedefaultseppunct}\relax
\EndOfBibitem
\bibitem[Kuzminykh \latin{et~al.}(2018)Kuzminykh, Polykovskiy, Kadurin,
  Zhebrak, Baskov, Nikolenko, Shayakhmetov, and Zhavoronkov]{cnn_rep}
Kuzminykh,~D.; Polykovskiy,~D.; Kadurin,~A.; Zhebrak,~A.; Baskov,~I.;
  Nikolenko,~S.; Shayakhmetov,~R.; Zhavoronkov,~A. 3D Molecular Representations
  Based on the Wave Transform for Convolutional Neural Networks.
  \emph{Molecular pharmaceutics} \textbf{2018}, \emph{15}, 4378--4385\relax
\mciteBstWouldAddEndPuncttrue
\mciteSetBstMidEndSepPunct{\mcitedefaultmidpunct}
{\mcitedefaultendpunct}{\mcitedefaultseppunct}\relax
\EndOfBibitem
\bibitem[Torng and Altman(2017)Torng, and Altman]{cnn_aa}
Torng,~W.; Altman,~R.~B. 3D deep convolutional neural networks for amino acid
  environment similarity analysis. \emph{BMC bioinformatics} \textbf{2017},
  \emph{18}, 302\relax
\mciteBstWouldAddEndPuncttrue
\mciteSetBstMidEndSepPunct{\mcitedefaultmidpunct}
{\mcitedefaultendpunct}{\mcitedefaultseppunct}\relax
\EndOfBibitem
\bibitem[Ryczko \latin{et~al.}(2018)Ryczko, Mills, Luchak, Homenick, and
  Tamblyn]{cnn_energy}
Ryczko,~K.; Mills,~K.; Luchak,~I.; Homenick,~C.; Tamblyn,~I. Convolutional
  neural networks for atomistic systems. \emph{Computational Materials Science}
  \textbf{2018}, \emph{149}, 134--142\relax
\mciteBstWouldAddEndPuncttrue
\mciteSetBstMidEndSepPunct{\mcitedefaultmidpunct}
{\mcitedefaultendpunct}{\mcitedefaultseppunct}\relax
\EndOfBibitem
\bibitem[He \latin{et~al.}(2016)He, Zhang, Ren, and Sun]{he2016}
He,~K.; Zhang,~X.; Ren,~S.; Sun,~J. Identity mappings in deep residual
  networks. European conference on computer vision. 2016; pp 630--645\relax
\mciteBstWouldAddEndPuncttrue
\mciteSetBstMidEndSepPunct{\mcitedefaultmidpunct}
{\mcitedefaultendpunct}{\mcitedefaultseppunct}\relax
\EndOfBibitem
\bibitem[Huang \latin{et~al.}(2017)Huang, Liu, Van Der~Maaten, and
  Weinberger]{densenet}
Huang,~G.; Liu,~Z.; Van Der~Maaten,~L.; Weinberger,~K.~Q. Densely connected
  convolutional networks. Proceedings of the IEEE conference on computer vision
  and pattern recognition. 2017; pp 4700--4708\relax
\mciteBstWouldAddEndPuncttrue
\mciteSetBstMidEndSepPunct{\mcitedefaultmidpunct}
{\mcitedefaultendpunct}{\mcitedefaultseppunct}\relax
\EndOfBibitem
\bibitem[Groom \latin{et~al.}(2016)Groom, Bruno, Lightfoot, and Ward]{CSD}
Groom,~C.~R.; Bruno,~I.~J.; Lightfoot,~M.~P.; Ward,~S.~C. {The Cambridge
  Structural Database}. \emph{Acta Crystallographica Section B} \textbf{2016},
  \emph{72}, 171--179\relax
\mciteBstWouldAddEndPuncttrue
\mciteSetBstMidEndSepPunct{\mcitedefaultmidpunct}
{\mcitedefaultendpunct}{\mcitedefaultseppunct}\relax
\EndOfBibitem
\bibitem[Ramachandran and Varoquaux(2011)Ramachandran, and Varoquaux]{mayavi}
Ramachandran,~P.; Varoquaux,~G. Mayavi: 3D visualization of scientific data.
  \emph{Computing in Science \& Engineering} \textbf{2011}, \emph{13},
  40--51\relax
\mciteBstWouldAddEndPuncttrue
\mciteSetBstMidEndSepPunct{\mcitedefaultmidpunct}
{\mcitedefaultendpunct}{\mcitedefaultseppunct}\relax
\EndOfBibitem
\bibitem[Stomberg \latin{et~al.}(1998)Stomberg, Li, Lundquist, and
  Albinsson]{cry_structure}
Stomberg,~R.; Li,~S.; Lundquist,~K.; Albinsson,~B. Investigation of three
  stilbene derivatives by X-ray crystallography and NMR spectroscopy.
  \emph{Acta Crystallographica Section C: Crystal Structure Communications}
  \textbf{1998}, \emph{54}, 1929--1934\relax
\mciteBstWouldAddEndPuncttrue
\mciteSetBstMidEndSepPunct{\mcitedefaultmidpunct}
{\mcitedefaultendpunct}{\mcitedefaultseppunct}\relax
\EndOfBibitem
\bibitem[Hartman \latin{et~al.}(2016)Hartman, Kudla, Day, Mueller, and
  Beran]{gipaw_pbe}
Hartman,~J.~D.; Kudla,~R.~A.; Day,~G.~M.; Mueller,~L.~J.; Beran,~G. J.~O.
  Benchmark fragment-based 1H{,} 13C{,} 15N and 17O chemical shift predictions
  in molecular crystals. \emph{Phys. Chem. Chem. Phys.} \textbf{2016},
  \emph{18}, 21686--21709\relax
\mciteBstWouldAddEndPuncttrue
\mciteSetBstMidEndSepPunct{\mcitedefaultmidpunct}
{\mcitedefaultendpunct}{\mcitedefaultseppunct}\relax
\EndOfBibitem
\bibitem[Chollet \latin{et~al.}(2015)Chollet, \latin{et~al.}
  others]{chollet2015keras}
others,, \latin{et~al.}  Keras. \url{https://keras.io}, 2015\relax
\mciteBstWouldAddEndPuncttrue
\mciteSetBstMidEndSepPunct{\mcitedefaultmidpunct}
{\mcitedefaultendpunct}{\mcitedefaultseppunct}\relax
\EndOfBibitem
\bibitem[Abadi \latin{et~al.}(2015)Abadi, Agarwal, Barham, Brevdo, Chen, Citro,
  Corrado, Davis, Dean, Devin, Ghemawat, Goodfellow, Harp, Irving, Isard, Jia,
  Jozefowicz, Kaiser, Kudlur, Levenberg, Man\'{e}, Monga, Moore, Murray, Olah,
  Schuster, Shlens, Steiner, Sutskever, Talwar, Tucker, Vanhoucke, Vasudevan,
  Vi\'{e}gas, Vinyals, Warden, Wattenberg, Wicke, Yu, and
  Zheng]{tensorflow2015-whitepaper}
Abadi,~M.; Agarwal,~A.; Barham,~P.; Brevdo,~E.; Chen,~Z.; Citro,~C.;
  Corrado,~G.~S.; Davis,~A.; Dean,~J.; Devin,~M. \latin{et~al.}  {TensorFlow}:
  Large-Scale Machine Learning on Heterogeneous Systems. 2015;
  \url{https://www.tensorflow.org/}, Software available from
  tensorflow.org\relax
\mciteBstWouldAddEndPuncttrue
\mciteSetBstMidEndSepPunct{\mcitedefaultmidpunct}
{\mcitedefaultendpunct}{\mcitedefaultseppunct}\relax
\EndOfBibitem
\bibitem[Ioffe and Szegedy(2015)Ioffe, and Szegedy]{BatchNorm}
Ioffe,~S.; Szegedy,~C. Batch normalization: Accelerating deep network training
  by reducing internal covariate shift. \emph{arXiv preprint arXiv:1502.03167}
  \textbf{2015}, \relax
\mciteBstWouldAddEndPunctfalse
\mciteSetBstMidEndSepPunct{\mcitedefaultmidpunct}
{}{\mcitedefaultseppunct}\relax
\EndOfBibitem
\bibitem[Nair and Hinton(2010)Nair, and Hinton]{RELU}
Nair,~V.; Hinton,~G.~E. Rectified linear units improve restricted boltzmann
  machines. Proceedings of the 27th international conference on machine
  learning (ICML-10). 2010; pp 807--814\relax
\mciteBstWouldAddEndPuncttrue
\mciteSetBstMidEndSepPunct{\mcitedefaultmidpunct}
{\mcitedefaultendpunct}{\mcitedefaultseppunct}\relax
\EndOfBibitem
\end{mcitethebibliography}
\providecommand{\latin}[1]{#1}
\makeatletter
\providecommand{\doi}
  {\begingroup\let\do\@makeother\dospecials
  \catcode`\{=1 \catcode`\}=2 \doi@aux}
\providecommand{\doi@aux}[1]{\endgroup\texttt{#1}}
\makeatother
\providecommand*\mcitethebibliography{\thebibliography}
\csname @ifundefined\endcsname{endmcitethebibliography}
  {\let\endmcitethebibliography\endthebibliography}{}

\end{document}